\documentstyle[aps,epsfig,subfigure]{revtex}
\begin{document}
\null\vskip0.5cm
\centerline{\Large\bf Space-like meson electromagnetic form factor}
\vskip0.5cm
\centerline{\Large\bf in a relativistic quark model} 
\vskip1.cm
\centerline{\Large L. Micu\footnote{E-mail: lmicu@theor1.theory.nipne.ro}}

\begin{center}
{Department of Theoretical Physics\\
Horia Hulubei National Institute of Physics and Nuclear Engineering\\
Bucharest POB MG-6, 76900 Romania}
\end{center}
\noindent
\begin{abstract}
The space-like electromagnetic form factor is expressed in terms of 
the overlap integral of the initial and final meson wave functions 
written as Lorentz covariant distributions of internal
momenta. The meson constituents are assumed to be
valence quarks and an effective vacuum-like field. The 
momentum of the latter represents a relativistic generalization of
the potential energy of the quark system.
 
The calculation is fully Lorentz covariant and the form factors
of the charged mesons are normalized to unity at $t$=0.
The numerical results have been obtained by freezing the
transverse degrees of freedom. We found  $r_\pi^2$=0.434 fm$^2$,
$r_{K^+}^2$=0.333 fm$^2$, $r_{K^0}^2$=-0.069 fm$^2$ by taking
$m_{u,d}$=430 MeV, $m_s$=700 MeV.

\end{abstract}
\vskip0.5cm
\section{Introduction}
In the standard model and the lowest order of perturbation the elastic
electron meson scattering is the result of one 
photon exchange between the electron and one of the quarks
in the meson. 
Formally the hadronic matrix element entering the expression of
the  scattering amplitude is parametrized in a model
independent Lorentz covariant way as: 
\begin{equation}\label{def}
<M(P')\vert~U_s(+\infty,0)~J_{em}^\mu(0)~U_s(0,-\infty)~
\vert M(P)>~=~{\cal T}^\mu~=~f_{em}(t)~(P+P')^\mu~
\end{equation}
\noindent
where $J_{em}^\mu$ is the elementary electromagnetic current
expressed in terms of free quark fields, $U_s(\tau,\tau')$ is the
time evolution operator of the system
due to quantum fluctuations of the colour field and
$t=(P-P')^2$  is a space-like momentum transfer. All the
information concerning the quark structure of the meson and the
effects of the quantum fluctuations of the coloured field is
contained in the electromagnetic form factor $f_{em}(t)$.
This makes the calculation of the electromagnetic form factor a rather
complicate problem. 

As shown in a careful analysis \cite{ils}, perturbative QCD
calculations solely cannot explain the observed pion and nucleon form
factors at presently accessible values of $t$ \cite{er}. The dominant
contribution seems to come from the hadronic wave functions which
are essentially nonperturbative. 

In fact this was to be expected from the uncertainty principle of
quantum mechanics. Point-like quarks and a real local interaction
between them could be observed only asymptotically. At low momentum
transfer only dressed quarks, the so-called constituent quarks, are
observable. Moreover, the electromagnetic current expressed in terms
of constituent quark fields is only approximately local and the
binding may also induce sizable effects.

In the present paper we calculate the form factors at low
momentum transfer, where mesons exhibit a
stable, time invariant structure in terms of constituent quarks. We
assume that, to a good extent, the electromagnetic current
expressed as a product of constituent quark fields may be considered
local. Consequently the matrix element (\ref{def}) will be related to
the  overlap integral between the wave function of the final meson and
that of the initial meson where one of the quarks has been replaced
by the quark interacting with the electromagnetic field. 
We therefore consider that the space-like form factor indicates to
what extent the initial system, where one of the quarks has been
replaced by the recoiling quark, represents the final meson. Similar
significance can be assigned also to the semileptonic form factors.

In this framework, at low space-like $t$ the form factor
$f_{em}(t)$ is the effect of a stable meson structure \cite{ils} and
one would expect it to decrease continuously
with $\vert t\vert$ because the larger is the momentum transfer, the
less probable is to find the recoiling quark in the final meson.
The decrease is then nothing else but the
consequence of the worse and worse matching of the initial with final
wave functions. 

Then the calculation of the form factors at low momentum transfer
amounts in finding a Lorentz covariant internal
wave function for the meson as a bound system made out of
independent constituents. 

One of the best known solution to this problem is the Bethe Salpeter
wave function. Completed with the expression of the dressed
quark-photon vertex satisfying the Dyson-Schwinger equation 
it led to remarkable results for low energy observables \cite{bs},
\cite{cr}. It must be noticed however that the formalism has 
ambiguities related to the definition of the relative quark momentum
and that the presence of negative energy states in the quark
propagators violates the real quark content of mesons.     
Also, in order to satisfy the usual normalization condition
of the form factor additional assumptions seem to be necessary
\cite{ils}, \cite{cr}.

Quark models also used solutions of
the Schr\"odinger equation with suitably chosen  potential terms.
The calculations have been performed at $t=0$ and  the decrease
with $\vert t\vert$ has been introduced from
phenomenological reasons \cite{isgw}.

Results independent of the specific form of the confining potential
have been obtained in the light-front approach \cite{ccp} where the
potential is formally included into the free mass term of
the quark-antiquark system. The form factors can be calculated at any
$t$, but manifest Lorentz covariance is lost because of the difference
between "good" and "bad" components of the current.

In this paper we present an original way to take into account the
confinement \cite{micu}.
The specific assumption of the model is that at low energy the
hadrons look like stable systems made of valence quarks {\it and} of
an effective vacuum-like field $\Phi$ representing countless
excitations of the quark-gluonic field which cannot be observed one
by one. The sum of all their momenta is the 4-momentum of the field
$\Phi$ which is supposed to be the relativistic generalization of the
potential energy of the quark system.

The main reason for introducing this effective description 
is suggested by the exemples taken from 
relativistic field theory, where a bound state is the result of an
infinite series of elementary interactions, but not the
instantaneous effect of an elementary process \cite{bs}. 

We therefore conjecture that the binding forces are the
result of a time average of elementary processes which cannot be
seen one by one if the time of observation is rather long (i.e.
if the momentum transfer from the projectile to the target is rather  
low).  We consider that the countless elementary excitations
continuously taking
place in a bound system deserve an average description in terms
of an effective field $\Phi$, not an individual, perturbative
treatment. Furthermore, the presence besides the valence quarks 
of an effective field $\Phi$ whose 4-momentum does 
not satisfy any mass constraint may also be useful for ensuring
a definite mass to the compound system. Indeed, as it is well
known, a system 
made of on-mass-shell particles having a continuous
distribution of relative momenta 
does not behave like a single particle because it does not
have a definite mass \cite{isgw}. An effective field may 
improve the situation by playing the role of a potential whose
contribution adds to the quark kinetic energy yielding a
definite value for the meson mass.

The main features of the model are presented in the next section.
We give the generic form of the meson state and calculate the
expression of the electromagnetic form factor by using the
canonical formalism of field theory.

The third section contains the numerical results obtained with
frozen transverse degrees of freedom.

In the last section we draw some conclusions concerning the
applicability of the model and comment on the values of the 
parameters yielding the best fit. 
%%%%%%%%%%%%%%%%%%%%%%%%%%%%%%%%%%%%%%%%%%%%%%%%%%%%%%%%%%%
\section{The model}

The main feature of the model is to suggest the generic wave
function of the bound quark system representing the meson. A
first requirement for the wave function is to be a Lorentz
covariant function of the independent quark coordinates because
the quarks really behave like independent particles in the
interaction with the electromagnetic field. It is obvious
that in order to satisfy this requirement
without introducing unphysical coordinates it is necessary to
work in momentum space. This will also help  
expressing easily the mass shell
constraints and the conservation laws. For this reason the
internal wave function of the meson will be given under the form
of the distribution function of the internal momenta.

Another constraint for the meson wave function follows from the
requirement to satisfy a single particle normalization condition.
Accordingly, the integral over the internal momenta must
converge, the relative momentum cut-off being provided by the
internal distribution function of the quark momenta. 

Recalling our conjecture that the binding potential can be
described with the aid of an effective field we assume that the
generic form of a single meson state is \cite{micu}:

\begin{eqnarray}
\label{meson}
&&\left.\vert M_i(P)\right\rangle
\int d^3p~{m_1\over e_p}
d^3q{m_2\over e_q}d^4Q \nonumber\\
&&\times\delta^{(4)}(p+q+Q-P)~\varphi(p,q;Q)
\times\bar u(p)\Gamma_M v(q)~\chi^+\lambda_i\psi~\Phi^\dagger(Q)~
\left.a^\dagger(p)b^\dagger(q) 
\vert 0 \right\rangle\,
\end{eqnarray}
where $a^+, b^+$ are the creation operators of the valence $q\bar q$
pair; $u,v$ are Dirac spinors and $\Gamma_M$ is a Dirac matrix 
ensuring the relativistic coupling of the quark spins. In the case
of pseudoscalar mesons $\Gamma_M=i\gamma_5$. The quark  creation and
annihilation operators satisfy canonical commutation relations and
commute with $\Phi^+(Q)$,  which represents the mean result of the
elementary quark-gluon excitations responsible for the binding.
Their total momentum $Q_\mu$ is not subject to any mass shell 
constraint and, in some sense, it is just what one needs to be 
added to the quark momenta in order to obtain the real meson 
momentum. This is in agreement with our assumption that $Q$ is the
relativistic generalization of the potential energy but we shall
avoid for a while making any definite assumptions about the
momentum carried by the effective field. 

The internal function of the meson is the Lorentz
invariant momentum distribution
function $\varphi(p,q;Q)$ which is supposed to be time independent, 
because it describes an equilibrium situation. This means that it
does not change under the action of internal strong forces and hence 
the time evolution operators $U_s(\tau,\tau')$ in eq. (\ref{def}) can 
be replaced by unity.
As mentioned above, the main r\^ole of $\varphi$ is to ensure
the single particle behaviour  
of the whole system, by cutting off the large relative momenta.
 
In the evaluation of the matrix element (\ref{def}) we shall use 
the cannonical commutation relations of the quark operators

\begin{equation}\label{cr}
\{a_i(k),a^\dagger_j(q)\}=\{b_i(k),b^\dagger_j(q)\}= (2\pi)^3
{e_k\over m}~\delta_{ij} \delta^{(3)}(k-q)
\end{equation}
and the expression of the vacuum expectation value of the effective
field which is defined as follows \cite{micu}:

\begin{equation}\label{vev}
\left\langle 0\right\vert~\Phi(Q_1)~\Phi^+(Q_2)~\left\vert 0 
\right\rangle~=
(VT_0)^{-1}~\int d^4~X~{\rm e}^{i~(Q_2-Q_1)_\mu~X^\mu}=
(2\pi)^4~(VT_0)^{-1}~\delta^{(4)}(Q_1-Q_2)
\end{equation}
where $V$ is the meson volume and $T_0$ is the 
characteristic time involved
in the definition of the mean field $\Phi$. $T_0$ is the time
needed by the bound state to be formed and hence we expect
it to be of the order of the hadronization time.

It is important to remark that the definition (\ref{vev}) is 
compatible with
the norm of the vacuum state if one takes $\Phi(0)=1$. We notice
also that the relation 
(\ref{vev}) has the character of a conservation law, just like the
commutation relations (\ref{cr}), both of them being 
necessary for the fullfilement of
the overall energy momentum conservation in the process.

As a first test of the model we evaluate the norm of the
single meson state (\ref{meson}) according to the usual procedure.
The exponent in the integral $\int_T~dX_0~{\rm
e}^{i(E(P)-E(P'))X_0}$ coming from the  
$\delta^{(4)}$ functions in eqs. (\ref{meson}) and (\ref{vev}) shall
be put equal to 0 and the integral equal to $T$, because the
uncertainty in the meson mass is expected to be much smaller than
$1/T$.

Observing that $T$ is the characteristic time involved 
in the definition of the effective field $\Phi$ in the case of a
moving meson, we write it as $T={E\over M}~T_0$ and get:  

\begin{equation}\label{norm}
\left\langle~{\cal M}(P')~\vert {\cal
M}(P)~\right\rangle~=
2E~(2\pi)^3~\delta^{(3)}(P-P')~{\cal J}
\end{equation}
where
\begin{equation}\label{J2} 
{\cal J}={-1\over 2MV}~\int d^3p~{m_1\over
e_p}~d^3q~{~m_2\over e_q}~d^4Q~
\delta^{(4)}(p+q+Q-P)\vert \varphi(p,q;Q)\vert^2~
Tr\left({\hat p+m_1\over 2m_1}~\gamma_5~
{\hat q-m_2\over2m_2}~\gamma_5\right)=1.
\end{equation}

This a remarkable result because it shows that the wave function of 
the many particle 
system representing the meson can be normalized like that of a single
particle if the integral $\cal J$ converges.

As a matter of consistency, we also remark the disappearance of the
rather arbitrary time constant $T_0$ from the expression
(\ref{norm}) of the norm.

We evaluate now the matrix element (\ref{def}) proceeding in the 
same manner as before. The expression of the 
electromagnetic current written in terms of free quark fields is
\begin{equation}\label{JEM}
J_{em}^\mu(x)={1\over(2\pi)^3}\sum_i~\kappa_i\bar\psi_i(x)
\gamma^\mu\psi_i(x)
\end{equation}
where $\kappa_i$ is the fraction of the proton charge carried by
the quark $i$. Introducing (\ref{JEM}) 
between the meson states (\ref{meson}) and using the relations
(\ref{vev}) and (\ref{cr}) to eliminate some integrals
over the internal momenta we obtain after a straightforward
calculation: 
\begin{equation}\label{tmu}
{\cal T}_\mu~={\cal T}_\mu^{(1)}+{\cal T}_\mu^{(2)}~=
-{1\over VT}~\int d^4Q~ {d^3p\over2e_p}
~{d^3q\over2e_q}~{d^3k\over2e_k}~\delta^{(4)}(p+q+Q-P)
~\varphi_i(p,q;Q)~
(t_\mu^{(1)}+t_\mu^{(2)})
\end{equation}
where
\begin{eqnarray}\label{t12}
t_\mu^{(1)}&=&\kappa_1\delta^{(4)}(k+q+Q-P')~\varphi_f^{(1)}(k,q;Q)
~Tr\left[\gamma_5(\hat k+m_1)\gamma_\mu(\hat p+m_1)
\gamma_5(-\hat q+m_2)\right]\\
t_\mu^{(2)}&=&\kappa_2~\delta^{(4)}(p+k+Q-P')~\varphi_f^{(2)}(p,k;Q)
~Tr\left[\gamma_5(-\hat k+m_2)\gamma_\mu(-\hat q+m_2)
\gamma_5(\hat p+m_1)\right].
\end{eqnarray}

The two terms in (\ref{tmu}) represent the contributions
of the valence quarks, $k$ is the momentum of the recoiling
quark, $P'$ is the final meson momentum
and $\varphi_{i,f}$ are the momentum distribution functions
of the initial and final mesons respectively. 

In the following we shall work in the Breit frame where
the momenta of the initial and final mesons are
$P=(E,~0,~0, P)$ and $P'=(E,~0,~0,-P)$ and the
electromagnetic form factor reads as: 

\begin{equation}\label{ff}
f_{em}(t)~=~{1\over\sqrt{4M^2-t}}~{\cal T}_0.
\end{equation}

In this frame it is an easy matter to show that $f_{em}(0)$=1 in
the case of a charged meson.
The demonstration makes use of  
$\delta^{(3)}(\vec{k}+\vec{q}+\vec{Q})$ to eliminate the
integrals over $\vec{k}$ 
in ${\cal T}_0^{(1)}$ and of the identity $(\hat p+m)
\gamma_0(\hat p+m)=2e_p~(\hat p+m)$ to reduce the number of 
projectors.  
Performing a similar operation on ${\cal T}_0^{(2)}$ and proceeding like
in the case of the norm, one gets
\begin{equation}
{\cal T}_0~=~2M~(\kappa_1 -\kappa_2) {\cal J}     
\end{equation}
which means 
\begin{equation}\label{qnorm}
f_{em}(0)=1
\end{equation}
if the meson wave function is properly normalized. 

In order to calculate the form factor at $t\ne 0$ we start by using
the
$\delta^{(3)}$ functions to eliminate the integrals over the momenta 
$\vec{q}$ and $\vec{k}$ in the expression of ${\cal T}_\mu^{(1)}$ and
over $\vec{p}$ and $\vec{k}$ in the expression of 
${\cal T}_\mu^{(2)}$.
After performing the traces over $\gamma$ matrices we get

\begin{eqnarray}\label{t1}
{\cal T}^{(1)}_\mu&=&{\kappa_1\over
VT}\int~de_p~dp_z~d\phi_p~d^4Q~{1\over4 e_k e_q}~
\delta(e_p+e_q+Q_0-E)~\delta(e_p-e_k)\nonumber\\
&\times&\varphi_i(p,q;Q)~\varphi_f^{(1)}(k,q;Q)
\{q_\mu~t+2\vec{P}\cdot\vec{Q}~(k_\mu-p_\mu)
+(k_\mu+p_\mu)[{\cal E}+{1\over2}t-2m_1(m_1+m_2)]\}
\end{eqnarray}
where 
\begin{equation}
{\cal E}=(E-Q_0)^2+\vec{P}^2-\vec{Q}^2+m_1^2-m_2^2.
\nonumber
\end{equation}
Next, by writing 
\begin{equation}
{1\over 2e_k}\delta(e_k-e_p)={1\over4P}\delta(p_z-P)
\end{equation}
and
\begin{equation}
{1\over2e_q}\delta(e_p+e_q+Q_0-E)=
{1\over2p_TQ_T}\delta\left(\cos\phi_p-{{\cal
E}-2e_p(E-Q_0)\over2p_TQ_T}\right) \end{equation}
we perform the integrals over $p_z$ and $\phi_p$ in eq. (\ref{t1}).

Then ${\cal T}^{(1)}_0$ becomes:

\begin{eqnarray}\label{final}
&&{\cal T}^{(1)}_0={\kappa_1\over 4VTP}~
\int d^4Q~\int_{e_{pm}}^{e_{pM}}de_p~\varphi_i(p,q;Q)~
\varphi_f^{(1)}(k,q;Q){1\over2p_T
Q_T\sqrt{1-\cos^2\phi_p}}\nonumber\\
&&\times\{2e_p\left[{\cal E}-2m_1(m_1+m_2)\right]+(E-Q_0)~t\}.
\end{eqnarray}

The integration limits over $e_p$ result from the kinematical
constraints $e_p^2\geq m_1^2+P^2$ and cos$^2\phi_p\leq 1$ 
which give:

\begin{eqnarray}\label{elim}
&&e_{pM}=
{(E-Q_0){\cal E}+
Q_T\sqrt{{\cal E}^2-4[(E-Q_0)^2-\vec{Q}_T^2](m_1^2+\vec{P}^2)}\over
2[(E-Q_0)^2-\vec{Q}_T^2]}\nonumber\\
&&e_{pm}={(E-Q_0){\cal E}-
Q_T\sqrt{{\cal E}^2-4[(E-Q_0)^2-\vec{Q}_T^2]
(m_1^2+\vec{P}^2)}\over2[(E-Q_0)^2-\vec{Q}_T^2]}.
\end{eqnarray}

The term ${\cal T}_0^{(2)}$ can be proccessed in the same manner, 
giving a similar expression.

Using the above results it is possible to calculate the
electromagnetic form factors at any any $t$, by
choosing an appropriate function $\varphi$.  In principle, the
calculation does not imply any other approximation, but it is hard to
believe that the multiple integral entering the expression of the
form  factor can be performed exactly.

The expression (\ref{final}) is, of course, valid for $t\ne0$, but the
infinite value one gets in the limit $t\to0$ seems to contradict
the normalization of the electric charge (\ref{qnorm}) which has been
demonstrated previously.

This is a disturbing situation which deserves a careful examination.
Looking back, we remark that the contradiction comes from the evaluation
of some $\delta$ functions:
\begin{equation}
\delta(\vec{p}+\vec{q}+\vec{Q}-\vec{P})\delta(e_p+e_q+Q_0-
E(P))\delta(\vec{k}+\vec{q}+\vec{Q}'-\vec{P}')\delta(e_k+
e_q+Q_0-E(P'))\delta^{(4)}(Q-Q')
\end{equation}
which have been written as
\begin{equation}\label{td0}
\delta(\vec{p}+\vec{q}+\vec{Q}-\vec{P})\delta(e_p+e_q+Q_0-
E(P))\delta(\vec{p}-\vec{k}-2\vec{P})\delta(e_p-e_k)
\delta^{(3)}(\vec{Q}-\vec{Q}')\delta(Q_0-Q'_0)
\end{equation}
at $t\ne$0, while at $t$=0 they have been written as
\begin{equation}\label{te0}
\delta(\vec{p}+\vec{q}+\vec{Q})\delta(e_p+e_q+Q_0-M)
\times\delta(\vec{p}-\vec{k})\delta(Q_0-Q'_0-M+M')\delta^{(3)}
(\vec{Q}-\vec{Q}'){1\over2\pi}\int {\rm e}^{i(M-M')X_0} dX_0.
\end{equation}
In the last expression the integral has been replaced by $T$ because
it was assumed  that the uncertainty in the meson mass is much
smaller than  $T^{-1}$.
As it was shown above, at $t=0$ the magnitude of $T$ does not really
matter because it disappears from the expression of the norm. 
This is not the case at $t\ne 0$ where $T$ remains in the final
expression of the form factor. Searching for a way to cure the
disagreement between the two cases we conjecture that $T$ must be
seen as the overlapping time of the initial and final systems. 
Of course, $T$ depends on the reference system we
consider. On the other hand, the form factor is scalar under Lorentz
transformations which means that we have to choose a particular
reference frame and write $T$ in terms of
Lorentz invariant quantities. (We recall that a similar problem
occurs in the definition of the scattering cross-sections, where
the flux of the incident particles is written in terms of the
relative momentum in the center of mass system.) We choose
the Breit frame and
consider that $T$ is the time necessary for an object of length
$L$ to pass along another object of length $L$. We put
then $T={L\over {P\over E}}$ where $L=L_0~{M\over E}$ is the
Lorentz contracted length of the meson box of size $L_0$ in its 
rest frame.

The numerical results quoted in this paper have been obtained 
by replacing the symmetry scheme based on the full Lorentz group
with the symmetry under the collinear group which is equivalent
with the flux tube model with frozen transverse degrees.
In this respect our working approximation may be considered as
opposite to the light front approach \cite{ccp} where the transverse
degrees of freedom are explicitly taken into account, while the
confining potential is assumed to give a fixed contribution to the
free meson mass. 

In the present approach the relativistic generalization of the
potential energy of the quark system is explicitly taken into
account as the 4-momentum of an effective field and the
mean contribution of the transverse degrees of freedom to the
meson energy is included in the quark masses.  The longitudinal and
the temporal degrees of  freedom are then the only active and the
multiple integral in eq.(\ref{final}) reduces to a simple one. 

In this case the expressions of ${\cal T}^{(1)}_0$ becomes:
\begin{eqnarray}\label{11t}
{\cal T}^{(1)}_0={2\kappa_1\over VT}~\int dQ_0~dQ_z~{dp_z\over
2e_p}~{dq_z\over 2e_q}~{dk_z\over
2e_k}~\delta^{(2)}(p+q+Q-P)\delta^{(2)}(k+q+Q-P')\nonumber\\
\times
\varphi_i(p,q;Q)\varphi_f^{(1)}(k,q;Q)\left\{(e_p+e_k)\left[(E-Q_0)^2-Q_z^2
-P^2-(m1-m2)^2\right]+e_q~t\right\}
\end{eqnarray}
Proceeding in the same manner as in the three dimensional case we
get: \begin{eqnarray}\label{delta}
&&{1\over 2e_k}={1\over 4P}\delta(p_z-P)\nonumber\\
&&{1\over 2e_q}\delta(e_p+e_q+Q_0-E)=\delta\left(m_2^2
+Q_z^2-(E-Q_0-e_p)^2\right).
\end{eqnarray}
Introducing now (\ref{delta}) in (\ref{11t}) and performing the
replacement $E-Q_0-e_1=m_2~x$ where $e_1=\sqrt{m_1^2+P^2}$ we obtain
finally: 
\begin{equation}\label{11fin}
{\cal T}^{(1)}_0={\kappa_1\over VT}~{m_1~m_2\over
P}~\int_1^\infty {1\over\sqrt{x^2-1}}\left({m_1\over
e_1}x+1\right) \varphi_i\varphi_f^{(1)}.
\end{equation}
In the above relation $\varphi_i,~\varphi_f^{(1)}$ are the
internal wave functions of the initial and final mesons and the
integration limits have been deduced  from the conditions
$Q_z^2\ge0$ or $e_2\ge m_2$ which make the square root real.

The expression of ${\cal T}^{(2)}_0$ can be obtained immediately
by interchanging $m_1$ with $m_2$ and then the space-like
electromagnetic form factor of pseudoscalar mesons reads:
\begin{equation}\label{11fem}
f_{em}(t)={m_1~m_2\over M^2~V_0~L_0}\int_1^\infty
{dx\over\sqrt{x^2-1}}\left[\kappa_1\left({m_1\over
e_1}x+1\right)\varphi_i\varphi_f^{(1)}+\kappa_{\bar 2}\left(
{m_2\over e_2}x+1\right)\varphi_i\varphi_f^{(2)}\right]
\end{equation}
where $e_{1,2}=\sqrt{m_{1,2}^2+P^2}$.
We remark that $V_0$ does not really appear at the end because
it will be cancelled out by a similar factor coming from the
normalization condition of the internal wave functions. This is
not the case for $L_0$, which still remains in the expression of
the electromagnetic form factor at $t\ne 0$. However, keeping in
mind that the form factor is really normalized to unity at
$t=0$, we shall take $L_0$ as a free parameter enabling us to
impose this constraint on the expression (\ref{11fem}).

A first consequence of the relation (\ref{11fem}), independent of the
particular form of the internal wave functions is that 
$f_{em}^{\pi^0,\eta,\eta'}(t)\equiv0$, 
while $f_{em}^{K^0,\tilde K^0}(t)\sim (m_s-m_d)$ and in principle
does not vanish.

%%%%%%%%%%%%%%%%%%%%%%%%%%%%%%%%%%%%%%%%%%%%%%%%%%%%%%%%%%%%%%%%%%

\section{Numerical results}

The numerical results we quote below have been obtained for
\begin{equation}\label{phi1}
\varphi(p,q;Q)={{\cal N}\over(p\cdot q)^n}={2^n{\cal
N}\over[(P-Q)^2-m_1^2-m_2^2]^n}
 \end{equation}
where ${\cal N}$ is a normalization constant. With the notations in
eq. (\ref{11fin}) one has
\begin{equation}\label{phi2}
\varphi_i~\varphi_{f}^{(1)}={{\cal
N}^2\over(m_1m_2)^{2n}\left(x^2+{P^2\over m_1^2}\right)^n}
\end{equation}

Observing that for $n=1,2$ the integral in the expression
(\ref{11fem}) can be performed analytically, we find:
\begin{eqnarray}\label{fem12}
f^{(1)}_{em}(t)&=&{{\cal N}^2\over M^2~V_0~L_0}
\sum_{i=1,\bar 2}\kappa_i~\left( 
{2\pi m_i^2\over  4m_i^2-t}+{2 m_i^2\over\sqrt{t^2-4m_i^2
t}}{\mathrm ln}{\sqrt{4m_i^2-t}+\sqrt{-t}\over \sqrt{4m_i^2-t}-\sqrt
{-t}}\right)\\
f^{(2)}_{em}(t)&=&{{\cal N}^2\over 
M^2~V_0~L_0}\sum_{i=1,\bar 2}\kappa_i~\left[
{4\pi m_i^2\over  (4m_i^2-t)}+{8 m_i^2\over
t^2-4m_i^2t}\right.\nonumber\\ &&+4\left({m_i^2\over
\sqrt{t^2-4m_i^2t}}
\left.+{m_i^4\over(4m_i^2-t)\sqrt{t^2-4m_i^2t}}\right)  {\mathrm
ln}{\sqrt{4m_i^2-t}+ \sqrt{-t}\over\sqrt{4m_i^2-t}-\sqrt{-t}}\right],
\end{eqnarray}   
which gives then: 
\begin{eqnarray}\label{r}
&&f^{(1)}_{em}(0)={{\cal N}^2\over M^2~V_0~L_0
}\left({\pi\over2}+1\right)\\ 
&&\left.{df^{(1)}_{em}(t)\over
dt}\right\vert_{t=0}={{\cal N}^2\over
M^2~V_0~L_0}\sum_{i=1,\bar 2}\kappa_i  {1\over m_i^2}
\left({\pi\over8}+{1\over6}\right)\\
&&{\mathrm lim}_{t\to-\infty}f^{(1)}(t)\sim{-{1\over
t}}{\mathrm ln}(-t )\\
&&f^{(2)}_{em}(0)={{\cal N}^2\over
M^2~V_0~L_0}\left({\pi\over4}+{2\over3}\right)\\
&&{df^{(2)}_{em}(t)\over dt}\left.\right\vert_{t=0}={{\cal N}^2\over
M^2~V_0~L_0}\sum_i\kappa_{i=1,\bar 2} {1\over m_i^2}
\left({\pi\over8}+{4\over15}\right)\\ &&{\displaystyle{\mathrm
lim}_{t\to-\infty}}f^{(2)}(t)\sim {1\over t^2}{\mathrm ln}(-t ).
\end{eqnarray}
The comparaison of the theoretical results with the experimental
data has been done by looking for the values of the parameter
$n$ and of the quark masses giving the lowest possible 
$\chi^2$, where:
\begin{equation}
\chi^2=\sum_i{(f_{th}^2(t_i)-f_{exp}^2(t_i))^2\over \sigma_i^2}.
\end{equation}
Here $\sigma_i$ are the experimental errors and the sum is
performed over all the experimental points quoted in \cite{amendolia}.
The values we found for the pion and kaon electromagnetic
form factors are given in Figs. 1 (a) and (b) respectively. They have been
obtained for $n$=1.5 and the quark masses $m_{u,d}$=430MeV,
$m_s$=700MeV. 
Our fit has $\chi_\pi^2$=69.4 for 45 experimental points and
respectively $\chi_K^2$=12.2 for 15 experimental points, while
the simple pole fit has $\chi_\pi^2$=52.5 and $\chi_K^2$=12. 
We also get $r_\pi^2$=0.434 fm$^2$, $r_{K^+}^2$=0.333 fm$^2$,
$r_{K^0}^2$=-0.069 fm$^2$
while the experimental values are
$(r_\pi^2)_{exp}$=0.431$\pm$0.01 fm$^2$, 
$(r_K^2)_{exp}$=0.34$\pm$0.05 and $(r_{K^0}^2)_{exp}=-0.054\pm$0.026
fm$^2$ \cite{amendolia}.
\begin{figure}[htb]
\begin{center}
\begin{tabular}{cc}
   \subfigure[\sf ]
{\epsfig{file=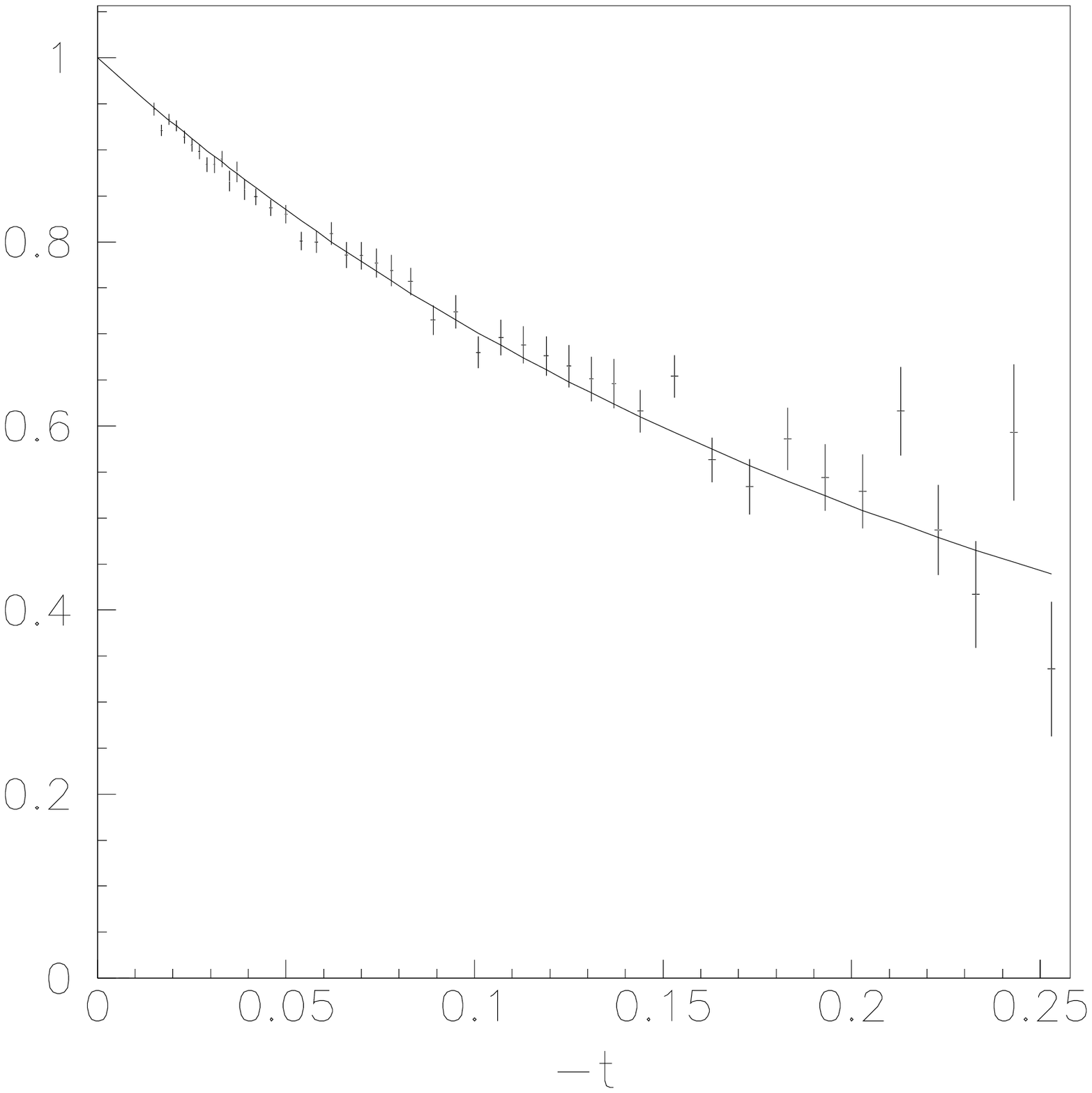,width=0.46\linewidth}}
   &
   \subfigure[\sf ]
{\epsfig{file=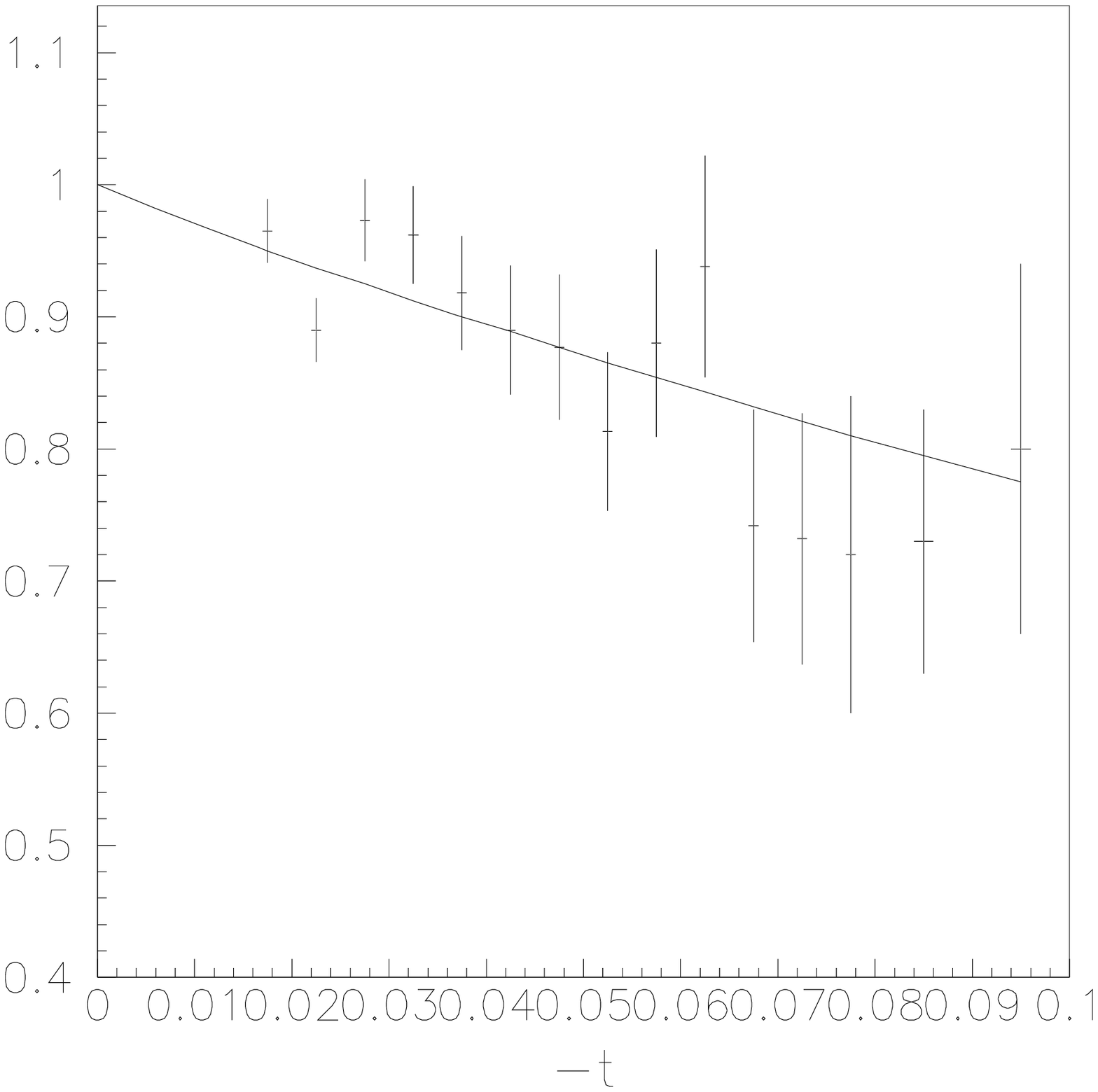,width=0.46\linewidth}}
\end{tabular}
\caption{ Comparison of the theoretical predictions for the square of
the pion (a) and kaon (b) electromagnetic form factors with data at
low $t$ (GeV$^2$/c$^2$). }  
\end{center}   \end{figure} 
We notice that the theoretical results are very sensitive
with respect to the up and down quark masses. This let us believe that
the numerical results quoted above may be improved by introducing a
difference between the lightest quarks. We notice also that the
values of the quark masses found from the best fit are a bit larger
than the currently used constituent masses \cite{ils,ccp,pd}.This is
normal taking into account that the contribution of the transverse
degrees  of freedom has been included in the quark masses. 

%%%%%%%%%%%%%%%%%%%%%%%%%%%%%%%%%%%%%%%%%%%%%%%%%%%%%%%%%%%%%%%%%
\section{Comments and conclusions} 

The model we presented in this paper is a relativistic model for
bound states. It enabled us to calculate the electromagnetic form
factors by means of the overlap integral over the internal
wave functions of the initial and final mesons. 

The internal function we used is just a suitable
trial function serving to illustrate the qualities of the model, not
the solution of a dynamical scheme. We remark that
in the infinite momentum limit it does not have any of the usual
asymptotic forms quoted in \cite{ils,ccp}. This is because the
distribution amplitude (\ref{phi1}) writes in terms of real
Lorentz scalar products which introduce "bad" components of momenta.
It demonstrates, however, that the existence of the meson form factors
can be explained through the stable, time invariant structure of the
mesons and also, that the colour fluctuations contributing to the
quark form factor can be neglected at low momentum transfer.

The remarkable point is that our results are unexpectedly consistent,
as it can be seen from the comparison of the pion radius with the box
length $L_0$. First we mention that $L_0$ cannot be infinite or
$0$. In the first case $f_{em}(t)=0$ for any $t\ne0$ and mesons would
behave like rigid bodies with respect to the  electromagnetic
interaction. In the second case mesons would be structureless. 

In our specific case $L_0$ can be calculated from the normalization
conditions of the meson state (\ref{norm}) and of the electromagnetic
form factor (\ref{qnorm}). Freezing two spatial degrees of freedom,
using the internal function (\ref{phi1}) and performing the
replacements $M-Q_0=(m_1+m_2)x$ and $Q_z=(m_1+m_2)y$ we get:

\begin{equation}\label{J1} 
{\cal J}={{\cal N}^2\over
2MV_0}~{2^n\over(m_1+m_2)^{2n-2}}\int_1^\infty dx~\int_{-\zeta}
^{\zeta}~dy~{1\over
[x^2-y^2-{m_1^2+m_2^2\over(m_1+m_2)^2}]^{2n}}
\sqrt{{x^2-y^2-1\over x^2-y^2-{(m_1-m_2)^2\over(m_1+m_2)^2}}}=1
\end{equation}  
where $\zeta=\sqrt{x^2-1}$ and the integration limits are
the consequence of reality of the norm. 

Then, by using the values of the parameters giving the best fit, we
obtain in the pion case $L_0\approx0.46$ fm, not far from the
pion radius which is roughly 0.65 fm. 

It is important to notice that the value we get for $L_0$ may
help us to put an upper limit for the time $T_0$ involved in
the definition of the effective field $\Phi$. To this end we recall
that $L_0$ is related with the overlapping time of the systems
representing the mesons and conjecture that
the  model is valid up to meson energies where this time is sensibly
larger than the time involved in the  definition of the effective
field $\Phi$. This means ${E~L_0\over P}>>{E~T_0\over M}$. Then, if
we expect the model holds at least up to $t=-1$ GeV$^2$ we get
$T_0<<0.7~10^{-25}$s. 

A last point we comment here concerns the continuity of the form
factor from the space-like to the time-like region. In principle our
expression (\ref{11fin}) can be continued at $t>0$, but it will have
no significance in the absence of a dynamical scheme of the strong 
interaction.
 
\vskip0.5cm 

{\bf Acknowledgements}
  The author thanks Prof. H. Leutwyler for the kind hospitality 
at the Institute of Theoretical Physics of the University of Bern
where this work was completed during author's visit supported by 
the Swiss National Science Foundation under Contract No. 7 IP 051219. 
The author thanks Fl. Stancu, University of Li\`ege, for useful 
suggestions and careful reading of the manuscript. 
The partial support of the Romanian Academy through the 
Grant No. 329/1997 is also acknowledged.


\vskip0.2cm
\noindent
\begin{thebibliography}{99}
\bibitem{ils}
N. Isgur and C. Llellwyn Smith, Phys. Rev. Lett. {\bf 52},1080
(1984); {\it ibid.} Nucl. Phys. {\bf B 317}, 526 (1989).

\bibitem{er}
A. V. Efremov and A. V. Radyushkin, Phys. Lett. {\bf B 94}, 254
(1980);  N. G. Stefanis, W. Schroers, H.-Ch. Kim, Phys. Lett. {\bf B
449}, 299 (1999); B. Meli\'c, B. Ni\v zi\'c, K. Passek, Phys. Rev.
{\bf D60}, 074004 (1999); V. M. Braun, A. Kodjamirian and M.
Maul, hep-ph/9907495.

\bibitem{bs}
H. Ito, W. W. Buck and F. Gross, Phys. Lett. {\bf 52}, 28 (1990);
{\it ibid.} {\bf B 287}, 23 (1992); {\it ibid.} {\bf B 351}, 24
(1996); M. Sawicki and L. Mankiewicz, Phys. Rev {\bf D 37}, 421
(1988); {\it ibid.} {\bf 40}, 3415 (1989).

\bibitem{cr}
C. D. Roberts, Nucl. Phys. {\bf A 605}, 475 (1996).

\bibitem{isgw}
B. Grinstein, M. B. Wise and N. Isgur, Phys. Rev. Lett. {\bf 56}, 298 
(1986); N. Isgur, D. Scora, B. Grinstein and M. B. Wise, Phys.
Rev. {\bf D 39}, 799 (1989).

\bibitem{ccp}
P. L. Chung, F. Coester and W. N. Polizou, Phys. Lett. {\bf B 205},
545 (1988); W. Jaus, Phys. Rev. {\bf D 41}, 3394 (1990); {\it
ibid.}, Phys. Rev. {\bf D 44}, 2851 (1991); F. Schlumpf, Phys. Rev.
{\bf D 50}, 6895 (1994); H.-M. Choi and C.-R. Ji, Nucl. Phys. {\bf A
618}, 291 (1997); {\it ibid.} Phys. Rev. {\bf D 59}, 034001
(1999); {\it ibid.} hep-ph 9908431.

\bibitem{micu} 
L. Micu,  Phys. Rev. {\bf D 55}, 4151 (1997). 

\bibitem{amendolia}
S. R. Amendolia et al. (NA7 coll.) Nucl. Phys. {\bf B 277}, 168
(1986); W. R. Molzen et al. Phys. Rev. Lett. {\bf 41}, 1213 (1978).

\bibitem{pd}
Particle Data Group, A. Manohar, Eur. Phys. J. C {\bf 3}, 337 (1998).


\end{thebibliography}
\end{document}